\documentclass[prl,reprint]{revtex4-1}
\usepackage{amsmath}  
\usepackage{amsfonts} 
\usepackage{graphicx} 
\usepackage{slashed} 

\newcommand{\be}{\begin{equation}}
\newcommand{\ee}{\end{equation}}
\newcommand{\wf}{wavefunction\;}
\newcommand{\wfs}{wavefunctions\;}
\newcommand{\infl}{influence function\;}
\newcommand{\infls}{influence functions\;}
\newcommand{\rmi}{\mathrm{i}}

\newcommand{\fy}{\slashed}

\begin{document}

\title{First quantized pair interactions, fermion loops and entanglement}

\author{A. F. Bennett}
\email{bennetan@oregonstate.edu} 
\affiliation{College of Earth, Ocean and Atmospheric Sciences\\
Oregon State University\\104 CEOAS Administration Building\\ Corvallis, OR 97331-5503, USA} 

\date{\today}

\begin{abstract}
Quantum Electrodynamics may be formulated as a Quantum Field Theory , and also as relativistic quantum mechanics by introduction of the Feynman-Stueckelberg parameter. As stated by M. Srednicki ({\it Quantum Field Theory}, Cambridge University Press, Cambridge and New York, 2007), ``\dots any relativistic quantum physics that can be treated in one formalism can also be treated in the other''. Entanglement of electrons at macroscopic spacelike separation has been observed (B. Hensen {\it et al.}, Nature, $\bf{526}$, 2015, doi:10.1038.nature/15759), yet according to Quantum Field Theory is restricted to separations of the order of the Compton wavelength $\hbar/m_e c$ (see e.g., S. J. Summers and R. Werner, Commun. Math. Phys. $\bf{110}$,247,1987). There is no restriction on entanglement in parametrized relativistic quantum mechanics  (A. F. Bennett, Ann. Phys. $\bf{345}$, 1, 2014). The parametrized formalism is extended here to pair annihilation and pair creation. The {\it ansatz} used in the formalism to develop the correct higher order corrections is justified here with plane wave summations. The conflict  over entanglement is further discussed  in terms of the Klein-Gordon propagator, and an invariant Bell's inequality is developed for spin -1/2.

\end{abstract}
\maketitle 

\section{Introduction}\label{S:intro}
The parametrization of relativistic quantum mechanics owes to both Feynman and  Stueckelberg. Feynman's development is reviewed  in \cite{Alv98}, while Stueckelberg's papers are reproduced in \cite{Stu_Works}. The formalism has been developed \cite{John69} to include the relativistic invariants, and as a textbook \cite{Fan93}. There has been little further attention. Srednicki \cite[p11]{Sr07} states that `` {\it quantum field theory}, the formalism on which position and time are both labels on operators, is much more convenient and efficient for most problems''. That statement is not disputed here. The issue is the shortcoming of QFT in regard to entanglement. As is well known, QFT is incapable of representing quantum entanglement. In the massive vacuum state, entanglement decays exponentially with spacelike separation measured in Compton wavelengths $\hbar/m_ec$\,. See e.g. \cite{Summ11}. The PM formalism provides a simple and unrestricted representation of entanglement in spacetime. The purpose of \cite{Benn2014} and this supplement is to establish the otherwise complete agreement between  parametrized relativistic quantum mechanics, hereafter PM, and QFT  as far as Quantum Electrodynamics. A further supplement will demonstrate parametrization of the Electroweak interaction.

 Interactions in PM are semiclassical, that is, electromagnetic interaction potentials are constructed classically from M\o ller currents \cite{BjDr64,Tay72,Benn2014}. The Bethe-Salpeter equation for bound states \cite{Salp51, Grei03} is obtained from the parametrized formalism without further conjecture.  Perturbation theory yields  the one-loop corections and axial anomalies of QFT.  The PM formalism lacks the causal commutator and anticommutator relations of QFT, but nevertheless yields ${\mathcal TPC}$ invariance and  the spin-statistics connections\cite{Benn2015,Benn2016}. Normal ordering of  Dyson series is natural, since the series involve integrals with respect to the parameter rather than to coordinate time.  The solution structure for the  parametrized Dirac equation is very close to that of the standard equation \cite{BjDr64,Grei03}, and comprehensive detail is available in  \cite{Benn2014}. 

The fundamental criticism of the standard Dirac formalism and of PM by implication must however be answered. If an electron cannot scatter into states of negative energy without lower bound, then the negative energy states must be identified as different particles, namely, positrons. How, then, in a particle--conserving formalism can pair annihilation such as 
\be\label{annih}
e^{-} + e^{+} \xrightarrow{t} \gamma+ \gamma
\ee
be represented? The symbol $t$ above the arrow indicates that coordinate time $x^0=t$ is increasing from left to right. It is shown here, by consideration of the parametrized  free influence functions,  that the ostensibly single-particle scattering amplitude obtained \cite{BjDr64} from  the Dirac equation is in fact an amplitude for the two-particle parametrized Dirac equation. Creation of a muon pair $\mu^-\mu^+$ subsequent to $e^-e^+$ annihilation is similarly shown to be a scattering event for the four-particle equation. The simple substitution in \cite{Benn2014}, which yielded the standard scattering corrections of higher order,  is justified here  by synthesizing internal lines from the initial and final free plane waves. Finally, a relativistic Bell's inequality is derived.

\section{The parametrized Dirac wave equation}\label{S:PDWE}

For a single spin--1/2 particle  the parametrized Dirac \wf is a four--spinor  $\psi(x,\tau)$. The event $x$ is in ${\mathbb R}^4$, while the parameter  $\tau$ is an independent variable in ${\mathbb R}$.  The event $x$ is also denoted by $x^\mu$ having indices $\mu=0,1,2,3$, with $x^0=ct$ where $c$ is the speed of light and $t$ is coordinate time. The Lorentz metric $g^{\mu\nu}$ on  ${\mathbb R}^4$ has  signature $(- + + +)$. The position $\bf{x}$  is denoted by $x^j$ having indices $j=1,2,3$. Thus $x=(ct,\bf{x})$.

The parametrized Dirac wave equation for $\psi$ is  
\be\label{PDIR}
\frac{\hbar}{\rmi c}\frac{\partial}{\partial \tau}\psi+\gamma^\mu\Big(\frac{\hbar}{\rmi}\frac{\partial}{\partial x^\mu}-\frac
{e}{c}A_\mu\Big)\psi=0
\ee
where $e$ is the charge of the particle, $c$ is the speed of light and $\hbar$ is the reduced Planck's constant. The $\gamma^\mu$ are the four Dirac matrices,  while the  Maxwell electromagnetic potential $A^\mu(x)$ is  independent of the parameter $\tau$.  The summation convention is assumed with respect to repetitions of Greek indices such as $\mu =0,1,2,3$\,. The covariant and contravariant indices $\mu,\nu,\dots$ will be omitted wherever convenient, as in $x= x^\mu$\,, $p = p^\mu$\, and  $p\cdot x = p^\mu x_\mu$ \,. Henceforth the units are chosen such that  $c=\hbar=1$.  The covariance of the theory with respect to the homogeneous Lorentz transformation $(x^\mu)' =\Lambda^{\mu}_{\;\;\nu}x^\nu$ and $ \psi'(x,\tau)=S(\Lambda)\psi(\Lambda^{-1}x,\tau)$ follows for $S(\Lambda)$  generated in the standard way  \cite{BjDr64}. No mass constant appears  in (\ref{PDIR}), but masses are introduced through boundary conditions as $\tau\to \pm \infty$. Feynman's development of QED using (\ref{PDIR}) has been reviewed by Garcia Alvarez  and Gaioli  \cite{Alv98}. The indefiniteness of the invariant bilinear form $\overline{\psi}\psi=\psi^\dag \gamma^0 \psi$ has impeded  \cite{Barut85, Evans98} the development of the parametrized Dirac formalism as a relativistic extension of quantum mechanics. 

\section{Plane wave solutions}\label{pws}
The plane \wf solutions of (\ref{PDIR}) in the absence of an electromagnetic field are of the form
\begin{subequations}\label{free}
\begin{gather}
{\bf f}^{(+)}_p(x,\tau)=\frac{{\bf u}_p}{(2\pi)^2}\exp[\rmi\chi^{(+)}]\,, \label{free1}\\ {\bf f}^{(-)}_p(x,\tau)=\frac{{\bf v}_p}{(2\pi)^2}\exp[\rmi\chi^{(-)}] \,.\label{free2}
\end{gather}
\end{subequations}
In (\ref{free}), $\chi^{(\pm)}=\chi^{(\pm)}(p,x,\tau)=p \cdot x \pm \varphi_p m_p \tau$,  $\varphi_p=\mathrm{sgn}(p^0)= p^0/E_p$ where $E_p=|p^0|$ is the energy of the particle, and $m_p=m(p)$ is the positive square root of $-p \cdot p$ for subluminal energy-momentum $p$. At constant phase $\chi^{(\pm)}$, $dx^0/d\tau=\pm m_p/E_p$ regardless of the value of $\varphi_p$ and so ${\bf f}^{(\pm)}_p$ propagates respectively forward $(+)$ and backward $(-)$ in   coordinate time $t=x^0$ as $\tau$ increases. The $4 \times 4$ block  of amplitudes $({\bf u}_p, {\bf v}_p)$ is a  basis of four  Dirac spinors constructed in the standard way \cite{Benn2014} from the elementary basis (the unit matrix) for the rest frame where ${\bf p}= {\bf 0}$\,. {\it The basis is an even function of} $p$\,. The orthonormality of the basis is expressed as $\overline{{\bf u}}_p{\bf u}_p = 1\,, \overline{{\bf v}}_p {\bf v}_p =-1, \overline{{\bf u}}_p {\bf v}_p  =\overline{{\bf v}}_p {\bf u}_p=0$\,. Antiparticle \wfs ${\bf h}^{(-)}_p$ are formed from particle \wfs ${\bf f}^{(+)}_p$ by combining the discrete symmetries \cite{Benn2014} of (\ref{PDIR}), yielding\footnote{The last equation in this string corrects an isolated error in Eq. (23) of Bennett (2014) \cite{Benn2014}.} ${\bf h}^{(-)}_p(x,\tau)=(\mathcal {TPC}\,{\bf f}^{(+)}_p)(-x,\tau)=-\rmi\gamma^5 {\bf f}^{(+)}_p(-x,\tau)=-\rmi {\bf f}^{(-)}_{-p}(x,\tau)$\,. Physically real particles and antiparticles are associated here with free on-mass-shell \wfs  having positive energy and positive mass, that is, they have energy $p^0>0$ and their \wfs are proportional to $\exp[+\rmi m_p \tau]$ with $m_p$ equal to the electron mass $m_e$\,. Thus a backward-propagating,  positive-energy  positron has the \wf that might otherwise be attributed to a `backward-propagating, negative-energy electron'.   While the coordinate time $t$ for a free positron \wf ${\bf h}^{(-)}_p(x,\tau)$ at constant phase does decrease as the parameter $\tau$ in increases, its spacelike coordinates $\bf x$ change in the same sense and at the same rates as those of its $\mathcal {TCP}$ conjugate \wf ${\bf f}^{(+)}_p(-x,\tau)$\,. 

It is readily verified that $\mathcal{TCP}$ conjugation commutes with not only the vector  interaction in (\ref{PDIR}), but with all the  invariant linear interactions (scalar, pseudoscalar, vector, axial vector and tensor) \cite{BjDr64}. In particular, the M\o ller currents and hence the semiclassical potentials transform as four-vectors.

\section{Free \infls}\label{props}
The free \infls $\Gamma^0_\pm (x'-x,\tau'-\tau)$ for (\ref{PDIR}) satisfy
\be\label{props}
\frac{1}{\rmi} \frac{\partial}{\partial \tau'} \Gamma^0_\pm + \frac{1}{\rmi} \gamma^\mu \frac{\partial}{\partial {x^\mu }'} \Gamma^0_\pm=\delta^4(x'-x)\delta(\tau'-\tau)\,.
\ee
Their Fourier transforms are
\begin{multline}\label{trans}
\Gamma^0_\pm(x,\tau)=\\ \left(\frac{1}{2\pi}\right)^5\int d^{\,4}dp \int d\varpi \Gamma^0_\pm(p,\varpi)\exp[\rmi(p \cdot x +\varpi \tau)]\,,
\end{multline}
where
\be
\Gamma^0_\pm(p,\varpi)=\frac{\varpi -\fy{p}}{\varpi^2-m_p^2\mp\rmi \epsilon}
\ee
where $\epsilon >0$ and $\fy{p}=\gamma^\mu p_\mu$\,. {\it Note that the subscripts} $\pm$ {\it on} $\Gamma^0_\pm$ {\it are not related to to the supercripts} $(\pm)$ {\it on} $f^{(\pm)}_p$. We have
\begin{multline}\label{G+}
\Gamma^{0}_+(x'-x,\tau'-\tau)=-\rmi \int d^4p\\ \times  \Bigg\{\theta (\tau'-\tau) \theta(p^0)-\theta(\tau-\tau')\theta(-p^0)\Bigg\}  \\ \times  \Bigg[{\bf f}^{(+)}_p(x',\tau')\overline{{\bf f}^{(+)}_p(x,\tau)}-{\bf h}^{(-)}_p(x',\tau')\overline{{\bf h}^{(-)}_p(x,\tau)} \Bigg]\,,
\end{multline}
and
\begin{multline}\label{G-}
\Gamma^{0}_-(x'-x,\tau'-\tau)=-\rmi \int d^4p \\ \times \Bigg\{\theta(\tau-\tau')\theta(p^0)-\theta (\tau'-\tau) \theta(-p^0) \Bigg\}  \\ \times  \Bigg[{\bf f}^{(+)}_p(x',\tau')\overline{{\bf f}^{(+)}_p(x,\tau)}-{\bf h}^{(-)}_p(x',\tau')\overline{{\bf h}^{(-)}_p(x,\tau)} \Bigg]\,,
\end{multline}
where $\theta(\tau)$ is the Heaviside unit step function. For  verbose descriptions of $\Gamma^0_\pm$, see \cite{Benn2014}.

\section{Pair annihilation}\label{S:pairann}
An incident  free \wf $\phi_i$ scatters off an electromagnetic field as a final  free \wf $\phi_f$\,, with scattering amplitude \cite{Tay72,Benn2014} 
\be\label {scamp}
S_{fi}=(\pm)\lim_{\tau \to + \infty}\int d^4x \;\overline{\phi_f(x,\tau)}(\omega_+ \phi_i)(x,\tau)
\ee
where $\omega_+$ is the forward M\o ller operator \cite{Benn2014} for  (\ref{PDIR}). Consistent with the normalization of the spinor basis, the leading sign in (\ref{scamp})  is $(+)$ for a final \wf that is forward-propagating  and $(-)$ for one that is backward-propagating. There is a nonvanishing amplitude for the mathematical contingency of a forward-propagating, positive-energy \wf scattering off two free photons and into a backward-propagating, negative-energy wavefunction. To leading order, the nontrivial and nonvanishing amplitude is 
\begin{multline}\label{oppie}
S_{fi}=-e^2\int d^5w'\int d^5w \\  \times \overline{{\bf h}^{(-)}_f(w')}\fy{A}(x')\Gamma^{0}_{+}(w'-w)\fy{B}(x){\bf f}^{(+)}_i(w)\,,
\end{multline}
where ${\bf h}^{(-)}_f={\bf h}^{(-)}_p$ for $p=p_f$, $w=(x,\tau)$, while the classical potentials $A^\mu(x)$ and $B^\mu(x)$ represent the two free photons, and $\Gamma^{0}_+$ as in (\ref{G+}) is the free forward \infl  for positive-energy electrons and protons \cite{Benn2014}\,. For clarity, the spins of the incident and final \wfs have not been stipulated and the additive amplitude with photons exchanged \cite{BjDr64} is suppressed. The integrals with respect to $\tau$ reduce to delta functions that enforce equality of the incident and final masses $m(p_i)$ and $m(p_f)$ respectively. Otherwise the parameterized scattering amplitude (\ref{oppie}) is the same as the standard Dirac amplitude \cite{BjDr64}. Identifying ${\bf h}^{(-)}_f$, the $\mathcal{TCP}$ conjugate of the final  `negative-energy electron' ${\bf f}^{(-)}_f$  as the parameter $\tau \to +\infty$\,, with a positive-energy positron injected as the coordinate time $t \to -\infty$ yields observed cross-sections for pair annihilation, now expressed as 
\be\label{annihtau}
e^{-}  +\gamma \xrightarrow{\tau} e^{+} + \gamma\,.
\ee 
In (\ref{annihtau}) it is the parameter $\tau$ which increases from left to right.  The standard first-quantized Dirac theory does not support more than one particle, much less the loss or gain of a particle. What becomes of the incident electron wave function in (\ref{oppie})? From whence came the final positron wave function? The issue is resolved by analyzing  the \infl or internal line in (\ref{oppie}) as in (\ref{G+}).

The second-order amplitude (\ref{oppie}) is thereby analyzed as a weighted sum of products of first-order amplitudes, for 

\begin{enumerate}
\item  a physical electron  ${\bf f}^{(+)}_i$ scattering into  a virtual (off mass shell) electron ${\bf f}^{(+)}_q$ of positive mass $m_q$\,, and a virtual electron ${\bf f}^{(+)}_p$ of positive mass $m_p$ scattering into a physical positron ${\bf h}^{(-)}_f$\,, or 

\item  a physical electron ${\bf f}^{(+)}_i$ scattering into a virtual positron ${\bf h}^{(-)}_q$ of positive mass $m_q$, and a virtual positron ${\bf h}^{(-)}_p$ of positive mass $m_p$ scattering into a physical positron  ${\bf h}^{(-)}_f$.  

\end{enumerate} 

In case (1) the virtual energies $p^0$ and $q^0$ are positive, while in case (2) they are negative. The minus sign preceding the second tensor product in (\ref{G+}) is consistent with the normalization of the spinor basis. The other two scattering contingencies $-{\bf f}^{(+)}_p \overline{{{\bf h}^{(-)}_q}'}$ and $+{\bf h}^{(-)}_p\overline{{{\bf f}^{(+)}_q}'}$ are kinematically excluded. Indicating  the virtual particles with braces $\{\dots\}$,  in case (i) above the interaction is now 
\be\label{annvirt}
e^{-} + \{e^{-}\} +\gamma \xrightarrow{\tau}  \{e^{-}\}+e^{+} +\gamma\,,
\ee
while in case (ii) it is 
\be\label{annvirt}
e^{-} + \{e^{+}\} +\gamma \xrightarrow{\tau} \{e^{+}\}+e^{+} +\gamma\,.
\ee
If the double integral with respect to $p$ and $q$ is uniformly weighted then the single-particle amplitudes vanish \footnote{This is most clearly seen by expanding  the single-particle amplitudes (\ref{scamp}) in the Born series.  Integrals over $\tau$ arise, and these may be postponed.} .  It is a textbook exercise \cite{BjDr64}, taking into account the kinematical constraints,  to verify that the amplitude products which are the  integrands on the right hand side of  (\ref{oppie}) modulo (\ref{G+}) are precisely those arising from the two-particle parametrized Dirac wave equation \cite{Benn2014}
 \be\label{TWO}
\frac{1}{\rmi}\frac{\partial}{\partial \tau}\Psi+\fy{D}(x)\otimes I_4\,\Psi+I_4\otimes \fy{D}(y)\,\Psi=0\,,
\ee
where $\Psi(x,y,\tau)$ is the two-particle wavefunction, while
\be\label{covDx}
D^\mu(x)=(1/\rmi)\partial /\partial x_\mu-e_1A^\mu(x)\,,
\ee
and
\be\label{covDy}
 D^\mu(y)=(1/\rmi)\partial /\partial y_\mu-e_2A^\mu(y)\,.
\ee 
Tensor products of the single-particle free \wfs  (\ref{free}) form a basis for the two-particle free wavefunctions. A nonseparable two-particle \wf provides an elementary and unrestricted representation of entanglement. The number of single-particle \wfs is conserved by the \textit{two}-particle scattering event (\ref{oppie}) analyzed with  (\ref{G+}). There are two incident wavefunctions, one physical and one virtual,  and there are two final wavefunctions, one virtual and one physical.

\section{pair creation}\label{S:prod}
It is observed that an electron and a positron can annihilate, followed by the creation of a muon and an antimuon. In the sense of $t$ increasing the interaction is
\be\label{teemupr}
e^{-} + e^{+} \xrightarrow{t}  \mu^{-} + \mu^{+}\,.
\ee    
The interaction is mediated by a virtual photon. In terms of $\tau$ increasing the interaction is
\be\label{taumupr}
e^{-} + \mu^{+} \xrightarrow{\tau} e^{+} + \mu^{-}\,.
\ee
The amplitude for the interaction may be constructed using (\ref{PDIR}) semi-classically. Consider the scattering of the antimuon into a muon.  The leading-order nontrivial amplitude is 

\be\label{am2m}
S_{f'i'}^{(1)}=\rmi e\int d\tau \int d^4 x \,\overline{f^{(+)}_{f'}(x,\tau)}\fy{A}(x)h^{(-)}_{i'}(x,\tau)\,.
\ee
The primed subscripts $f'$ and $i'$ indicate that the \wfs have the energy-momenta of the incident antimuon and final muon respectively.
The scattering potential $A^\lambda$ is the (virtual) photon owing to the electron-positron current. That is, 
\be\label{Max}
A^{\lambda}(x)=e\int d^4y\, D^{\lambda\nu}(x-y) J_{\nu}(y)
\ee
where $D^{\lambda\nu}$ is the standard \infl for a massless vector boson \cite{Zee03}, while $J^{\nu}$ is the M\o ller current owing to the electron and positron. To leading order,
\be\label{e2p} 
J^{\lambda}(y)=\int d\sigma\,  \overline{h^{(-)}_{f}(y,\sigma)}\gamma^{\lambda}f^{(+)}_{i}(y,\sigma)\,.
\ee
Note that the current has been concatenated \cite{ArHoLa83,Benn2014} with respect to the parameter, that is, integrated for all $\sigma$\,.  The scattering amplitudes include  spurious factor $2\pi\delta(o)$, which is interpreted as a a large parameter value  $T$. It can be suppressed by  defining the concatenation, as it arises in the Lagrangian \cite{Benn2014}, to be 
\be\label {catL}
\lim_{T \to \infty}\frac{1}{T}\int_{T/2}^{-T/2}\bigg(\dots \bigg)d\tau\,.
\ee

The amplitude (\ref{am2m})--(\ref{e2p}) again raises the issue for first-quantized formalism: a positive-energy physical antimuon is being scattered into a different particle, namely, a positive-energy physical muon. There is of course an identical amplitude for the physical electron being scattered into a physical positron by the potential  sustained by the antimuon-muon current. The issue is resolved in two stages. First, the boson \infl is analyzed as a sum of products. Second, the number  of vertices is doubled from two to four by introducing two trivial vertices. At each vertex there is an on-shell \wf for a physical particle and an off-shell \wf for a virtual particle. There is also a virtual photon at each of the two original vertices, but none at either of the two trivial vertices. The amplitude is then precisely of the form arising from the four-particle parametrized Dirac equation. Some details now follow. 

First, the concatenation in (\ref{Max}) is postponed, and the massless vector boson \infl  is replaced with the \infl for a massive vector boson \cite{Zee03}. The spectral representation of the \infl becomes
\be\label{mbos}
D^{\lambda\nu}(k,\varpi)=\frac{G^{\lambda\nu}(k,\varpi)}{k\cdot k+\varpi^2}\,,
\ee
where $k$ is the four-vector wavenumber, while $\varpi$ is the frequency with respect to the parameter $\tau$ as in $\exp(\rmi \varpi \tau)$\,. The numerator is variously expressed as
\be\label{numer}
G^{\lambda\nu}(k,\varpi)=g^{\lambda\nu}+\frac{k^\lambda k^\nu}{\varpi^2}=\sum_{j=1}^3 \varepsilon_j^\lambda \varepsilon_j^\nu\,,
\ee
where $\varepsilon_j^{\lambda}(k)$ is a polarization amplitude such that $k_\mu\varepsilon_j^{\mu}=0$ for $j=1,2,3$\,. It is evident that the boson \infl is an integral of products of virtual, $\tau$-dependent `preMaxwell' photons \cite{AhHo06}, or plane waves that are off the `shell' where $k\cdot k =-\varpi^2$\,.  
Second, the incident antimuon in (\ref{am2m}) is expressed as 
\begin{multline}\label{virtvex}
h^{(-)}_{i'}(x,\tau)=\frac{1}{\rmi}\int d^4z \{\theta(\tau-\rho)\Gamma^0_+(x-z,\tau-\rho)-\\ \theta(\rho-\tau)\Gamma^0_-(x-z,\tau-\rho)\}h^{(-)}_{i'}(z,\rho)\,.
\end{multline}
The fermion influence function $\Gamma^0_{-}$ has \cite{Benn2014} an  analysis  similar  to (\ref{G+}). The expression (\ref{virtvex}) can accordingly be interpreted as the trivial scattering of the incident physical antimuon $h^{(-)}_{i'}$ into a virtual particle. The other virtual particle (in the tensor product within either \infl) is incident to the physical vertex at $(x,\tau)$\,. The incident virtual particle scatters off the virtual photon and into the final physical muon $f^{(+)}_{f'}$\,. 

The analysis of the boson influence function,  the introduction of trivial vertices and the analysis of the fermion \infl convert both the $e^-e^+$ side and the $\mu^-\mu^+$ side of the single-particle {\it ansatz } (\ref{am2m})--(\ref{e2p}) to instances of (\ref{scamp}). Then (\ref{am2m})--(\ref{e2p}) is  expressed as an amplitude for a multiple-particle parametrized Dirac equation, without any physical particle being converted by scattering into a different physical particle. The expression is verbose, but the {\it ansatz} is equivalent and provides a  convenient short-hand. 

\section{Loops}\label{Bubb}

Higher-order corrections agree with QFT. It suffices, by way of explanation, to consider a second-order scattering of a single particle. For example, a free off-shell electron $\psi_i$ scatters off photon $A_\mu$ and another photon $B_\mu$ , into a free off-shell electron $\psi_f$. The scattering amplitude  is  

 \begin{multline}\label{scatt}
 S_{fi}^{(2)}= e^2\int d{\,^5}w'  \int d{\,^5}w\, \rmi\, \overline{{\psi_f(x',\tau')} }\fy{A}(x')\Gamma^{0}_+(w'-w) \\   \times \fy{B}(x) {\psi_i(x,\tau)}\,,
 \end{multline}
 where $w=(x,\tau)$. The invariant measures in the integrals are $d^{\,5}w=d\tau d^{\,4} x$, etc.  The integrand may  be expressed  as the  trace 
\begin{multline}\label{tracesactt}
\mathrm { Tr}  \left[ \,\fy{B}(x)\Gamma_{-}^0(w-w')\fy{A}(x')  (-\rmi)\, \psi_f(x',\tau') \overline{\psi_i(x,\tau)} \right] ^*.
\end{multline}
The loop is completed by making the substitution
\be\label{subsG}
\rmi \psi_f(x',\tau')\overline{\psi_i(x,\tau)} \to  \Gamma^0_+ (w'-w).
\ee
The substitution is a nmemonic for the plane wave expansion of  a free influence function, as is  now shown. 

Let the initial and final \wfs be $\psi_i(x,\tau)={\bf f}^{(+)}_i(x,\tau)$ and $\psi_f(x',\tau')={\bf f}^{(+)}_f(x',\tau')$, where subscripts $i$ and $f$ indicate the energy -momenta $p_i$ and $p_f$ respectively.
The trace (\ref{tracesactt}) becomes
\be\label{traceS}
\mathrm{Tr}\left[\,\fy{B}(x)\Gamma_{-}^0(w-w')\fy{A}(x') (- \rmi) \,{\bf f} ^{(+)}_f(x',\tau')\overline{{\bf f}_i^{(+}) (x,\tau)}\, \right]^*.
\ee 
The traces are now summed.
\begin{enumerate}
\item For $\tau'>\tau$, add the scattering amplitudes for all $p_i$ and for all $p_f$  which have positive energies,  with weight $\delta (p_f-p_i)$. 
\item For $\tau'<\tau$, subtract the scattering amplitudes for all $p_i$ and for $p_f$ which have negative energies, also with weight $\delta(p_i-p_f)$.  
\item Repeat the preceding two steps with the positron scattering amplitude, in which case  trace is  the trace is
\begin{multline}\label{traceP}
\mathrm{Tr}\left[\,\fy{B}(x)\Gamma_{-}^0(w-w')\fy{A}(x') \,(- \rmi) \,{\bf h} ^{(-)}_f(x,\tau)\overline{{\bf h}_i^{(-)} (x',\tau')}\, \right]^*,
\end{multline} 
where the positron \wfs are
\begin{multline} \label{posi}
{\bf h}^{(-)}_p(x,\tau)=({\mathcal TCP}\, {\bf f}_p^{(+)})(-x,\tau) \\ =-\rmi \gamma^5 {\bf f}_p^{(+)}(-x,\tau)=-\rmi {\bf f}_{-p}^{(-)}(x,\tau)\,.
\end{multline} 
\end{enumerate}
The grand sum of the outer products of the wavefunctions is  $-\rmi \Gamma_{+}^0(w-w')$, see \cite{Benn2014}. Further rearrangement yields
\begin{multline}\label{fur}
S^{(2)}_{fi}= \\ -e^2\int'\int \mathrm{Tr}\left[\fy{B}(x)\Gamma_+^{0}(w'-w)\Gamma_-^{0}(w-w')\fy{A}(x')\right]\,.
\end{multline}
The line $\Gamma^0_+(w-w')$ in the scattering amplitude connects the vertices in the diagram for the scattering of an electron off two photons, and so completes the fermion loop. Similar constructions yield the self-mass correction, the vertex correction and the axial anomaly \cite{Benn2014}. The further and standard additional minus sign is introduced whenever the substitution yields a closed fermion loop \cite[p151]{BjDr64}, \cite[p316]{Sr07}. 

\section {entanglement in algebraic Quantum Field Theory}\label{S:alg}

The algebraic QFT of entanglement has attracted wide interest. See for example  \cite{Haag58,Araki62,Schlie65,Summ11}. The principal finding, possibly first reported in \cite{Araki62}, is most clearly expressed in \cite{Summ87c} as follows. Consider a a pair of spacetime events in a quantum vacuum. The  measure $\beta$ of entanglement decreases essentially as $\beta(Y)=1+(\sqrt{2}-1)\exp(-Y/\lambda)$ where $\lambda=1/m$  is the Compton wavelength, and  $Y=|y^0|-|{\bf y}|$  the pure timelike translation from $y$ to the nearest lightlike path. See \cite{Summ11} for more precise estimates of $\beta$\,. 

The principle is basic. Relativistic quantum  field theory is founded on the Klein-Gordon equation
\be\label{KG}
\partial_\mu \partial ^\mu \Phi +m^2\Phi=0,
\ee
where $ \Phi(x)$ is a scalar quantum field acting on Fock space.  
The vacuum-to-vacuum amplitude or influence function $G(x'-x)$, which satisfies
\be\label{KGinfl}
{\partial_{\mu} }'{\partial^{\mu }}'G(x'-x)+m^2G(x'-x)=-\delta^4(x'-x)\,,
\ee
has the explicit forms \cite[p30]{Hua04}
\be\label{KGTL}
G(x'-x)=\frac{m}{8\pi\sqrt{-u}}H^{(2)}_1(m\sqrt{-u})
\ee
for $u=(x'-x)\cdot(x'-x)<0$, that is, for timelike separation, and
\be\label{KGSL}
G(x'-x)=-\frac{\rmi m} {4\pi^2\sqrt{u}}K_1(m\sqrt{u})
\ee
for $u>0$, that is, for spacelike separation. There is a singularity in $G$ in the neighborhood of the light cone, such that
\be\label{KGLC}
\lim_{u \to 0} G(u)=\frac{-1}{4\pi}\delta(u)\,.
\ee
As  $|z| \to \infty$ the Hankel function $H_1^{(2)}(z) \sim \mathcal {O}(|z|^{-1})$, and the modified Bessel function $K_1(z) \sim \mathcal{O}(\exp(-|z|))$. Thus there can  is long-range timelike entanglement, but spacelike entanglement is restricted to separation at the scale of the Compton wavelength $1/m$\,. Massless particles will be addressed in a sequel article.

\section{spin-1/2 entanglement}\label{S:ente}

Consider now entanglement in parametrized relativistic quantum mechanics. The development of a covariant Bell's inequality closely paraphrases the nonrelativistic development for Pauli spinors \cite{WeiQM13}. The PM wave equation  for two fermions has the form (\ref{TWO}), where the gauge bosons in $\fy{D}$ are sustained by the sums of the current for each of the two fermions. That is, the pair is interacting. Of course, (\ref{TWO}) also defines the dynamics for non-interacting free fermions considered here.  While parametrization is not involved in the derivation of the inequality, it is essential to the first-quantized formulation of the singlet.

 Let $A$ and $B$ be two Stern-Gerlach apparatus having a common rest frame. Let $a\,,b$ and  $\,c$ be unit four vectors that are pure spacelike in the  frame of the apparatus, and that coincide with three choices for the directions of the static magnetic fields in the apparatus. That is, and ignoring transposition,  $a=(0,\Hat{\boldsymbol {a}})$ where $a\cdot a = -\boldsymbol{\Hat{a}}\boldsymbol{\cdot}\boldsymbol{\Hat{a}}=-1$\, and similarly for $b$ and $c$.  The spacelike components of these covariant vectors in any other frame are not those of  magnetic fields. The latter transform instead as in \cite[\S11.10]{Jack99}. Let $p= (E_p\,, \boldsymbol {p})$ and $q= (E_p\,,- \boldsymbol {p})$ be the energy-momenta of oppositely moving free Dirac fermions that  enter $A$ and $B$ respectively.  Assume further that $a\cdot p=b\cdot p = c \cdot p =0$\,. It follows that $a\cdot q= b\cdot q = c \cdot q =0$\,. The eigenvalues of the spin operators $\gamma^5\fy{a}, \gamma^5\fy{b}
$ and $\gamma^5\fy{c}$ are therefore good quantum numbers for both fermions. Those quantum numbers are all $\pm1$. For example, let the eigenstates of $\gamma^5\fy{a}$ be $\psi_{\uparrow}$ and $\psi_{\downarrow}$ where 
\be\label{updown}
\gamma^5\fy{a}\psi_\uparrow=+\psi_\uparrow\;\mathrm{and}\; \gamma^5\fy{a}\psi_\downarrow=-\psi_\downarrow\,.
\ee
Both fermions are assumed to have the same positive energy $E_p$ where  $E_p=\sqrt{m^2 + \boldsymbol {p}\boldsymbol{\cdot} \boldsymbol {p}}$, \, so $\overline{\psi_\uparrow}\psi_\uparrow=\overline{\psi_\downarrow}\psi_\downarrow=1$ while \,$\overline{\psi_\uparrow}\psi_\downarrow=\overline{\psi_\downarrow}\psi_\uparrow=0$\,.  Finally,  assume  the two identical free fermions   form a spin 1/2 singlet with respect to the same  spin 1/2 operator $\gamma^5\fy{a}$\,.  That is, the two-particle \wf is necessarily
\be\label{twoferm}
\Theta(x,y,\tau)=\frac{1}{\sqrt{2}}\Bigl(\psi_\uparrow(x,\tau)\otimes\psi_\downarrow(y,\tau)-\psi_\downarrow(x,\tau)\otimes\psi_\uparrow(y,\tau)\Bigr)\,.
\ee
Tensors such as (\ref{twoferm}) may be construct from un-parametrized Dirac wavefunctions. However there is no  single-time wave equation for the tensor.  It is the parameter $\tau$ in (\ref{TWO}) which synchronizes the two fermions.

The state (\ref{updown}) is a singlet since
\be\label{sing}
\Bigl(\gamma^5\fy{a}\otimes\mathbf{1}+\mathbf{1}\otimes\gamma^5\fy{a}\Bigr)\Theta(x,y,\tau)=0\,.
\ee
Again, the parametrized, first quantized representation (\ref{twoferm}) and the two-particle wave equation  (\ref{TWO}) make clear that covariant integration of a multiple particle state with respect to $\tau$ is possible, and in effect with respect to $x^0$ and $y^0$.  The energies of the two particles may differ, and so  the local coordinate times  may not advance at the same rate. 

If correlations of spin measurements owe to a local hidden variable $\lambda$, then \cite{WeiQM13}
\be\label{hidden}
\beta \equiv |\langle \gamma^5\fy{a}\otimes\gamma^5 \fy{b}\rangle_\lambda -\langle \gamma^5\fy{a}\otimes \gamma^5 \fy{c}\rangle_\lambda| -\langle \gamma^5\fy{b}\otimes\gamma^5 \fy{c}\rangle_\lambda\leq  1 \,.
\ee
 Now choose  $a^\mu,\,b^\mu$ and  $c^\mu$ so that
\be\label{choose}
 a\cdot b =0 \;\mathrm{and}\;  c=(a+b)/\sqrt{2}\;.
\ee
Without loss of generality let $\boldsymbol{\Hat{a}}=(1,0,0)\,,\,\boldsymbol{\Hat{b}}=(0,1,0)$, and so $\boldsymbol{p}=(0,0,p^3)$\,. That is, the particle beams are perpendicular to the static magnetic fields. Splitting of the beams owes to the gradient of the strength of the static magnetic field \cite{WeiQM13}, but that is not a consideration here. It is readily shown that 
\be\label{qmb}
\overline{\psi_\uparrow}\gamma^5\fy{b}\psi_\uparrow=\overline{\psi_\downarrow}\gamma^5\fy{b}\psi_\downarrow=0\, 
\ee
hence quantum mechanical correlations with respect to (\ref{twoferm}) satisfy
\be\label{qmcorr1}
\langle \gamma^5\fy{a}\otimes\gamma^5 \fy{b}\rangle_{QM}=\langle \gamma^5\fy{b}\otimes\gamma^5 \fy{a}\rangle_{QM}=0\,,
\ee
and 
\be\label{qmcorr2}
\langle \gamma^5\fy{a}\otimes\gamma^5 \fy{a}\rangle_{QM}=\langle \gamma^5\fy{b}\otimes\gamma^5 \fy{b}\rangle_{QM}= a \cdot a =  b\cdot b= -1\,.
\ee
Note that free fermions are all $\propto \exp(\rmi \theta \tau)$, thus the PM correlations in (\ref{qmcorr1}) and (\ref{qmcorr2}) are independent of $\tau$. It follows from those equations that
\begin{multline}\label{qmabc1}
|\langle \gamma^5\fy{a}\otimes\gamma^5 \fy{b}\rangle_{QM} -\langle \gamma^5\fy{a}\otimes \gamma^5 \fy{c}\rangle_{QM}| \\- \langle \gamma^5\fy{b}\otimes\gamma^5 \fy{c}\rangle_{QM}=  \sqrt{2}\,,
\end{multline}
which contradicts (\ref{hidden})\,. A two-particle state is ``entangled" if $\beta \ge 1$, and ``maximally entangled" if $\beta = \sqrt{2}$\,.

While the free fermions in (\ref{twoferm}) are $\propto \exp(\rmi \theta)$,  the factors $\psi_\uparrow(x)$ and $\psi_\downarrow(x)$ can be localized in spacetime. That is, unrestricted entanglement in coordinate time $x^0$ is admitted.

The fundamental difference between the representations of entanglement in QFT and PM may be exposed as follows.
The two-point correlation in QFT is the influence function or propagator $G(x'-x)$, which may be expressed as \cite[p56]{Sr07} 
\be\label{Tord}
G(x'-x)=\rmi<0|T\Phi(x')\Phi(x)|0>
\ee
where $T$ is the time-ordering operator. The field $\Phi(x)$ in (\ref{Tord}) is any normalized free field $\Phi$ for which the dispersion relation is $p\cdot p+m^2=0$. The propagator  is formally the analog of the plane wave expansions  \cite[(29)-(34)]{Benn2014}. Indeed, the information in the propagator is exactly the information in the plane wave expansions.  The correlation in PM is the expectation of a tensor product of operators as in (\ref{qmabc1}), relative to the two-particle wave function (\ref{twoferm}). The supports of the single-particle wave functions such as $\psi_\uparrow(x,\tau)$ and $\psi_\downarrow(y,\tau)$  are disjoint if the separation $x-y$ is spacelike and if  $(x-y)\cdot (x-y) \gg 1/m$.

\section{Summary}\label{S:summ}

The pre-eminance of Quantum Field Theory is unlikely to be exceeded soon by any alternative formulation of relativistic quantum physics. While parametrized relativistic quantum mechanics reproduces every finding of Quantum Field Theory, the former does lack the flexibility of involving only a single field.  PM has the advantage of a realistic representation of relativistic entanglement. 

\section*{References}
\bibliography{paraDirac_Bib}

\end{document}